\documentclass[aps, pre,twocolumn,superscriptaddress]{revtex4-1}
\usepackage{graphicx}% Include figure files
\usepackage{dcolumn}% Align table columns on decimal point
\usepackage{bm}% bold math
\usepackage{amssymb,amsfonts,amsmath} 
\usepackage[usenames,dvipsnames]{xcolor}
\begin{document}

\preprint{APS/123-QED}

\title{Interplay between Kinetics and Dynamics of Liquid-Liquid Phase Separation in a Protein Solution Revealed by Coherent X-ray Spectroscopy}% Force line breaks with \\

\author{Anastasia Ragulskaya}
\email{anastasia.ragulskaya@uni-tuebingen.de}
\affiliation{Institut\,f\"{u}r\,Angewandte\,Physik,\,Universit\"{a}t\,T\"{u}bingen,\,Auf\,der\,Morgenstelle\,10,\,72076\,T\"{u}bingen,\,Germany}
\author{Nafisa Begam}
\affiliation{Institut\,f\"{u}r\,Angewandte\,Physik,\,Universit\"{a}t\,T\"{u}bingen,\,Auf\,der\,Morgenstelle\,10,\,72076\,T\"{u}bingen,\,Germany}
\author{Anita Girelli}
\affiliation{Institut\,f\"{u}r\,Angewandte\,Physik,\,Universit\"{a}t\,T\"{u}bingen,\,Auf\,der\,Morgenstelle\,10,\,72076\,T\"{u}bingen,\,Germany}
\author{Hendrik Rahmann}
\affiliation{Department Physik, Universit\"{a}t Siegen, Emmy-Noether-Campus, Walter-Flex-Strasse 3, 57072 Siegen, Germany}
\author{Mario Reiser}
\affiliation{Department Physik, Universit\"{a}t Siegen, Emmy-Noether-Campus, Walter-Flex-Strasse 3, 57072 Siegen, Germany}
\affiliation{European X-ray Free-Electron Laser GmbH, Holzkoppel 4, 22869 Schenefeld, Germany}
\author{Fabian Westermeier}
\affiliation{Deutsches Elektronen-Synchrotron (DESY), Notkestrasse 85, 22607 Hamburg, Germany}
\author{Michael Sprung}
\affiliation{Deutsches Elektronen-Synchrotron (DESY), Notkestrasse 85, 22607 Hamburg, Germany}
\author{Fajun Zhang}
\email{fajun.zhang@uni-tuebingen.de}
\affiliation{Institut\,f\"{u}r\,Angewandte\,Physik,\,Universit\"{a}t\,T\"{u}bingen,\,Auf\,der\,Morgenstelle\,10,\,72076\,T\"{u}bingen,\,Germany}
\author{Christian Gutt}
\email{christian.gutt@uni-siegen.de}
\affiliation{Department Physik, Universit\"{a}t Siegen, Emmy-Noether-Campus, Walter-Flex-Strasse 3, 57072 Siegen, Germany}
\author{Frank Schreiber}
\email{frank.schreiber@uni-tuebingen.de}
\affiliation{Institut\,f\"{u}r\,Angewandte\,Physik,\,Universit\"{a}t\,T\"{u}bingen,\,Auf\,der\,Morgenstelle\,10,\,72076\,T\"{u}bingen,\,Germany}

\date{\today}% It is always \today, today,
\begin{abstract}
Microscopic dynamics of complex fluids in the early stage of spinodal decomposition (SD) is strongly intertwined with the kinetics of structural evolution, which makes a quantitative characterization challenging. In this work, we use x-ray photon correlation spectroscopy to study the dynamics and kinetics of a protein solution undergoing liquid-liquid phase separation (LLPS). We demonstrate that in the early stage of SD, the structural relaxation kinetics is up to 40 times slower than the dynamics and thus can be decoupled. The kinetic decay rate is inversely proportional to time in the early stage, followed by a nearly constant behavior during the coarsening stage. The microscopic dynamics can be well described by hyper-diffusive ballistic motions with a relaxation time exponentially growing with time in the early stage followed by a power-law increase with fluctuations. These experimental results are further supported by simulations based on the Cahn-Hilliard equation. 
The established framework is applicable to other condensed matter and biological systems undergoing phase transitions and may also inspire further theoretical work.

\end{abstract}

%\keywords{XPCS}%Use showkeys class option if keyword
\maketitle

%\tableofcontents

Liquid-liquid phase separation (LLPS) is a fundamental process that has significant consequences in many biological studies, e.g. investigations of protein crystallization \cite{Durbin}, bio-materials and diseases \cite{Gunton2007}, as well as for food \cite{Gibaud} or pharmaceutical industry \cite{Raut}. In cells, LLPS can play a crucial role in driving functional compartmentalization without the need for a membrane and as a ubiquitous cellular organization principle implicated in many biological processes ranging from gene expression to cell division \cite{Shin2017,Boeynaems2018,Hondele2020}. LLPS is a complex process involving structural evolution of the new phases, microscopic dynamics of density fluctuation, and the thermal fluctuation of domain interfaces, as well as global diffusive motion. While both the kinetics of structural evolution and the microscopic dynamics are important for the formation and properties of the various condensates, research so far mainly focused on the structural growth kinetics and the dynamics remains largely unknown.

Dynamics of a system in equilibrium can be described using the fluctuation of the static structure factor, $S(q)$, where the average density is constant and $\rho(t)$ fluctuates around this mean value. In this case, $\overline{S(q)}$ is constant and dynamics can be directly extracted.
For a system undergoing phase separation, the interplay between dynamics and kinetics may lead to a more complex scenario, i.e. the temporal average density (and accordingly, $\overline{S(q)}$) changes and fluctuates at the same time. 
%.
At the late stage of LLPS via SD,  theoretical studies have shown that the domain coarsening follows dynamic scaling, i.e. during structural evolution, the domains remain essentially statistically self-similar at all times \cite{FURUKAWA}. As a result, $\overline{S(q)}$ does not change if the measurements are performed at length scales corresponding to the average domain size \cite{Brown, Brown1999}. Thus, in principal, the time-dependent fluctuations of $S(q)$ can be described by the two-time intensity covariance, which equals the square of the two-time structure factor under the "Gaussian decoupling approximation" \cite{Brown, Brown1999}. Therefore, domain coarsening dynamics can be accessed by calculating the speckle-intensity covariance \cite{Brown, Brown1999}. 

However, the early stage of LLPS is much more complex because the time-dependent $S(q,t)$ cannot be described by dynamic scaling. The interplay between kinetics and dynamics makes a quantitative description of the microscopic dynamics challenging. So far, such a theory to predict or an experimental method to probe the dynamics in the early stage of SD is missing. Nevertheless, the dynamic property of LLPS is a vital input to predict the dynamic properties of the resulting condensates in biological systems. Studies of LLPS in cells have shown that the exact biological function of the resulting condensates strongly depends on their dynamic properties, such as a dense liquid, a gel or a glassy state \cite{Shin2017}. Therefore, detection in the early stage of the condensates are needed, which requires a characterization of the dynamics of LLPS. Furthermore, as LLPS in cells is generally triggered by a gradient of constitutes instead of a temperature quench, the dynamic properties of LLPS also reflect the material transport in the crowded environment of a multi-component system.

Despite the recent progress in the field of phase separation \cite{Shin2017, Berry_2018}, the interplay between the kinetics and dynamics at the early stage remains elusive. %
What is the influence of the change of $S(q)$ with time on dynamics? Whether one can distinguish the kinetic and dynamic contribution experimentally? What are the main features of dynamics in the early stage of LLPS? In principle, both kinetics and dynamics of LLPS can be simultaneously accessed experimentally using X-ray photon correlation spectroscopy (XPCS). The XPCS investigation of the fluctuations during the late, domain coarsening, stage was demonstrated vividly by the work of Malik et al. in a sodium borosilicate glass \cite{Malik1998}. Microscopic dynamics of systems undergoing phase transition displays rich non-equilibrium behavior on the length-scales ranging from micrometer to single-protein size on time scales from hundred of seconds down to microseconds \cite{Grimaldo}. XPCS has been used in different soft condensed matter systems \cite{Lurio,Perakis,Lumma,Bandyopadhyay}. However, due to experimental difficulties in working with beam-sensitive samples \cite{Jeffries2015,Fluerasu2008,Ruta2017}, XPCS of protein-based systems became possible only recently with a special design of experimental procedure   \cite{Begam2021,Girelli,Gutt,Vodnala2016}.

In this Letter, we apply state-of-the-art XPCS in the ultra-small angle scattering (USAXS) geometry to investigate the structural evolution and microscopic dynamics during LLPS in aqueous protein solutions. The system consists of the globular protein bovine serum albumin (BSA) in the presence of trivalent salt yttrium chloride ($\textrm{YCl}_3$) \cite{DaVela16,Matsarskaia_2016_JPhysChemB}. The phase behavior of this system has been established to exhibit a lower critical solution temperature (LCST) phase behavior \cite{Matsarskaia_2016_JPhysChemB, Braun2018, DaVela16}. The kinetics of LLPS of this system has been well studied using USAXS \cite{DaVela16,DaVela20}. Here we aim to explore both the growth kinetics and the microscopic dynamics in order to distinguish their role in the early stage of LLPS and the domain coarsening using XPCS in the USAXS geometry. Combined with 2D Cahn-Hilliard simulations, we demonstrate that the dynamic information can be distinguished from the growth kinetics with a factor of $\sim 40$ difference in time scales.
The method established in this work can be used to access the dynamics during phase transitions in a broad range of soft matter and biological systems.

BSA and $\textrm{YCl}_3$ were purchased from Sigma-Aldrich. Sample preparation followed our previous work \cite{DaVela16,DaVela20}. The stock solutions of protein and salt were mixed at 21~$^\circ$C with an initial protein concentration of 175 mg mL$^{-1}$ and salt concentration of 42 mM. The obtained solution was equilibrated, centrifuged and the dense phase was used for further experiments.

XPCS experiments were performed in USAXS mode at PETRA III beamline P10 (DESY, Hamburg, Germany) at an incident X-ray energy of 8.54 keV ($\lambda$ = 1.452 \r{A}) and beam size of $100$ $\mu m$ $\times$ 100 $\mu m$. The sample-to-detector distance was 21.2 m, which corresponds to a $q$ range of $3.2\times10^{-3}$ nm$^{-1}$ to $3.35\times10^{-2}$ nm$^{-1}$, where $q=4\pi /\lambda\cdot \sin \theta$ and $2\theta$ is the scattering angle. The data were collected by an EIGER X 4M CCD detector with $75$ $\mu m$ $\times$ 75 $\mu m$ pixel size. The sample was filled into capillaries of 1.5 mm in diameter. Details can be found in the supplementary information (SI). 

\begin{figure}%[h]
 \centering
\includegraphics{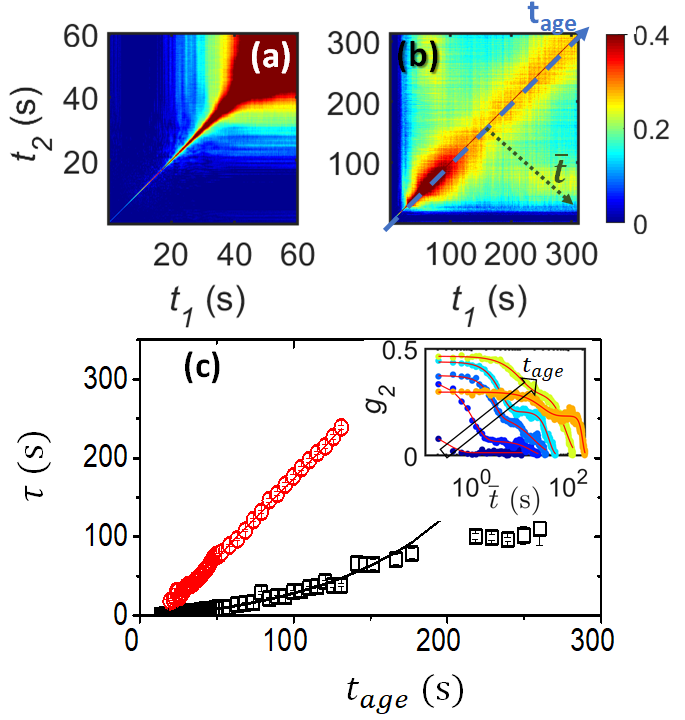} 
 \caption{Two relaxation modes revealed by XPCS. \textbf{(a)} TTC for the first 60s and \textbf{(b)} 312s after a temperature jump from 10 $^\circ$C to 40 $^\circ$C for 312 s experiment at  $q$=4.4$\mu m^{-1}$. The colorbar corresponds to the $G(q,t_1,t_2)$ values. The slow mode describes the sudden decay in correlation in \textbf {(b)} around 20-30s, exhibiting a square feature in TTC. Blue long-dash line shows the direction of the increase of experimental time $t_{age}$, and the short-dash line displays the direction of the $g_2$-cuts. The inset of \textbf{(c)} shows $g_2$ obtained from \textbf{(b)}.  \textbf{(c)} Relaxation time $\tau$ evolution as a function of $t_{age}$ for fast (black squares) and slow (red circles) modes at $q$=4.4$\mu m^{-1}$. $\tau$ of the fast mode  follows an exponential growth (black solid line) before 200s, whereas $\tau$ for the slow mode grows linearly with time (red solid line).}
 \label{fgr:TTC}
\end{figure}

The time-resolved 2D speckle patterns collected in XPCS measurements are analyzed using a two-time correlation function (TTC) $G\left(q, t_1, t_2\right)$ \cite{Grubel}:
\begin{eqnarray}\label{eq:TTC}
 G(q,t_1,t_2)=\frac{\overline{I(t_1)I(t_2)}-\overline{I(t_1)}\cdot\overline{I(t_2)}}{[\overline{I^2(t_1)}-\overline{I(t_1)}^2]^{\frac{1}{2}}\cdot[\overline{I^2(t_2)}-\overline{I(t_2)}^2]^{\frac{1}{2}}},
\end{eqnarray}
where the average is over pixels with the same momentum transfer $q \pm \Delta q$. Here $t_1$ and $t_2$ are the times at which the intensity correlation is calculated. 
Typical results of TTC are presented in Fig. \ref{fgr:TTC}. In the TTC of 60s, a relaxation signal appears a few seconds after the temperature jump, initially with a fast decay rate, but quickly broadening. The TTC of 312s shows that the main relaxation along the diagonal turns into a steady state after the quick broadening. A new slow relaxation mode appears, and its correlation shows a sudden decay around 20-30s, leading to a square feature in the TTC.

The correlation function, $g_2(q, t_{age})$, as a function of time is determined along the cuts perpendicular to the diagonal ($t_1=t_2$) of the TTC (the direction is marked by the green dotted line in Fig. \ref{fgr:TTC} (b)) for each specific experimental time $t_{age}=(t_1+t_2)/2$. $g_2(q, t_{age})$ functions can be fitted using the Kohlrausch-Williams-Watts (KWW) relation to determine the characteristic relaxation time ($\tau$) and the shape parameter ($\gamma$) \cite{Sinha} as functions of $q$ and $t_{age}$. As we can see in the inset of Fig. \ref{fgr:TTC} (c), the $g_2(q,t_{age}, \overline{t})$ functions display two-decay relaxations - a fast and a slow mode, which can be modelled using two exponential decay functions:
\begin{eqnarray} \label{eq:g2two}
g_2=\beta_1\exp(-2\left(\frac{\overline{t}}{\tau_1}\right)^{\gamma_1})+\beta_2\exp(-2\left(\frac{\overline{t}}{\tau_2}\right)^{\gamma_2}),
\end{eqnarray}
where the index 1 corresponds to the fast component and the index 2 to the slow component of the relaxation. The parameters $\tau$ and $\gamma$ are functions of both $q$ and $t_{age}$.
Here $\overline{t}=t_2-t_1$ is a delay time, $t_{age}$ is the time that passed from the start of the heating, $\beta(q)$ is the speckle contrast, which can vary from $0$ (incoherent scattering) to $1$ (fully coherent scattering) for ergodic processes. The $g_2$ functions were averaged over a small range of time to increase the signal-to-noise ratio (SI Table S1).

Relaxation times, $\tau_1$ and $\tau_2$, extracted from the KWW fits are shown in Fig. \ref{fgr:TTC} (c). The fast mode exhibits two distinct stages of the evolution of relaxation time: initially, $\tau$ increases exponentially ($t_{age}<150$ s), followed by a modulation around 100 s in the later stage of the experiment (Fig. \ref{fgr:TTC} (c)). Similar two-stage behavior has been reported in XPCS investigations of Wigner glasses \cite{ANGELINI}, glassy ferrofluids \cite{Robert}, and egg-white \cite{Begam2021}  and has been suggested as a general feature near the glass transition. 

The slow mode corresponds to the broad square-like feature observed on the TTC in Fig. \ref{fgr:TTC} (b). The relaxation time of the slow mode linearly increases as a function of $t_{age}$. The shape parameter $\gamma_2$ of the slow mode shows a jump in the early stage from values less than 2 to higher ones and then fluctuates at values around 2.7 (Fig. S2 (a)). Such high values indicate that there are different phenomena taking place in the early and later stages and these stages hardly correlate with each other. Based on a comparison (SI Fig. S3) of the behavior of the $\overline{S(q)}$ for the experiment and simulation (discussed later)  and matching with the real-space picture (obtained from simulation, SI Fig. S4), it seems that the appearance of the slow mode correlates with the transition from the density fluctuation to the coarsening.

In the early stage of LLPS, due to the strong coupling between structural growth and domain fluctuation, the resulting relaxation rates (Fig. \ref{fgr:TTC} (c)) may have both contributions. Dynamical scaling cannot describe the time-dependent scattering profiles in the early stage as expected (SI Fig. S1). In order to distinguish or decouple the contributions, we determine and compare the kinetic relaxation rate with the XPCS results.  

\begin{figure}

  \includegraphics[width=6cm]{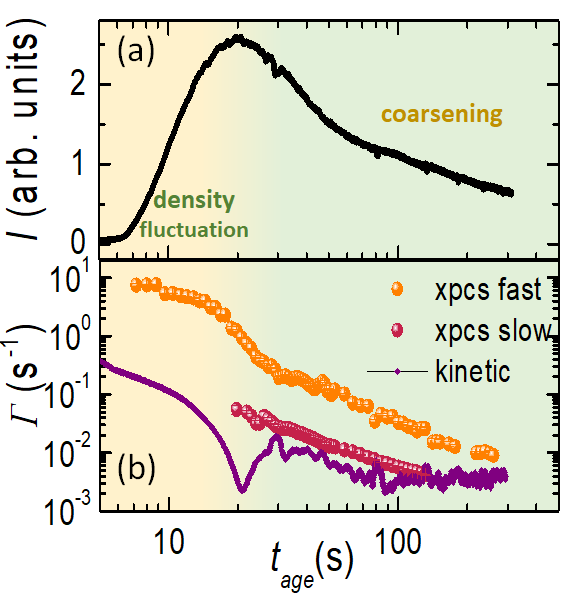}
  \caption{Decoupling between kinetics and dynamics. \textbf{(a)} Structure factor evolution with $t_{age}$ at $q$=4.4$\mu m^{-1}$. \textbf{(b)} Comparison of relaxation rate for fast mode of XPCS (orange), for slow mode of XPCS (coral red) with kinetic relaxation rate (purple, obtained from USAXS) for $q$=4.4$\mu m^{-1}$. Yellow and green regions in \textbf{(a)} and \textbf{(b)} correspond to the density fluctuation and domain coarsening stages of LLPS, respectively.} 
  \label{fgr:relaxation_rate}
\end{figure}

The evolution of a kinetic rate $\Gamma_{kinetic}$ at a specific $q$ can be calculated through the changes in the scattering intensity $I(q,t_{age})$ \cite{Kurt}:
\begin{equation}\label{eq:rate}
{\Gamma}_{kinetic}(q,t_{age})= \frac{1}{2I(q,t_{age})}\frac{dI(q,t_{age})}{dt_{age}}.
\end{equation}
The scattering intensity $I(q,t_{age})$ was calculated for each value of $t_{age}$ based on USAXS experiments as the mean intensity $I(q \pm \Delta q,t_{age})$ of all pixels inside the ring $q \pm \Delta q$ (Fig. \ref{fgr:relaxation_rate} (a)), but only the absolute values of $dI/dt_{age}$ are used for comparison. 

Fig. \ref{fgr:relaxation_rate} (a) shows the evolution of $S(q)$ with $t_{age}$ at $q$=4.4$\mu m^{-1}$. The corresponding kinetic decay rate calculated using Eq.\ref{eq:rate} shown in Fig. \ref{fgr:relaxation_rate} (b) exhibits a quick decrease with aging time initially and becomes nearly a constant after about 100s. For comparison, the decay rates of both slow and fast modes from XPCS measurements at $q=4.4\mu m^{-1}$ are also shown in Fig. \ref{fgr:relaxation_rate}.  It is interesting to see that the slow-mode decay rate is similar to the kinetics during the early stage of coarsening. 

The fast mode starts at $t_{age} \sim$ 7 s, before which the dynamics is too fast to be detected with the current XPCS measurements (limited by values of exposure and delay times). 
The decay rate is up to 40 times higher than the kinetic relaxation in the early stage and about one order of magnitude faster in most of the coarsening stage (up to $\sim 150s$). Thus, these two effects are decoupled, and the fast mode corresponds to the microscopic dynamics of domain fluctuations. 

\begin{figure}

  \includegraphics[width=8cm]{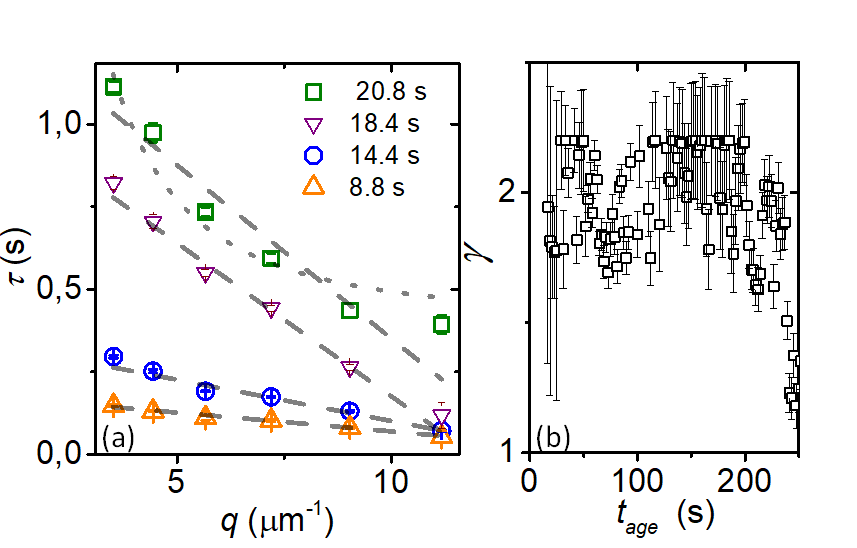}
  \caption{Hyper-diffusive dynamics revealed by XPCS:   \textbf{(a)} $q$-dependent of $\tau$ and  \textbf{(b)} shape parameter $\gamma$ as a function of $t_{age}$ for the fast mode dynamics of XPCS. Grey lines represent fit with $\tau \sim q^{-1}$ (dashed) and $\sim q^{-2}$ (dotted).} 
  \label{fgr:relaxation_time_vs_q_and_KWW}
\end{figure}

Now we can interpret the fast mode as a reflection of the principal dynamics of the system during LLPS. The $q$-dependency of its relaxation time is shown in Fig. \ref{fgr:relaxation_time_vs_q_and_KWW} (a). In the early stage, $t_{age}<20s$, $\tau$ follows an inverse linear relationship with $q$, $\tau \sim q^{-1}$. With increasing time, it becomes non-linear with $\tau \sim$ between $q^{-1}$ and $q^{-2}$. In the mean time, the shape parameter $\gamma>1$ (Fig. \ref{fgr:relaxation_time_vs_q_and_KWW} (b). These results demonstrate the domain fluctuation dynamics to follow the hyper-diffusive ballistic motion in the early stage.

 As discussed in the Introduction, during the domain coarsening step, due to the dynamic scaling effect, the covariance of the TTC is directly connected with the fluctuation of $S(q)$, i.e. the dynamics of domain fluctuation. Here we perform simulations to verify if the kinetic relaxation rate can be decoupled with the dynamics in this stage as well. A system undergoing spinodal decomposition can be described by the Cahn-Hilliard equation (CH)  \cite{Cahn1958,Cahn1959}, which gives the concentration at each point of the simulated map with time. After rescaling of parameters  \cite{Sappelt} and adding a temperature jump \cite{Sciortino}, we can write:

\begin{equation} \label{eq:CH}
\frac{\partial u(\mathbf{r}, t)}{\partial t}=\nabla\left[m(u) \nabla\left(-\frac{T_c-T}{T_c}u+u^{3}-\nabla^{2} u\right)\right],
\end{equation}
where $u(\textbf{r},t)$ - rescaled time-dependent local concentration, $m(u)$ - mobility function, $T_c$ - critical temperature. Further details of the simulation can be found in the SI.

Based on the simulations, the evolution of the 2D field of concentration was calculated (SI Fig. S4 (c)-(f)), and the corresponding scattered speckle pattern $I(q,t_{age})$ was calculated (i.e. image in reciprocal space) for each time step as a square of the magnitude of the 2D fast Fourier transform of the fluctuations of the concentration \cite{Barton}. The dynamics of the simulated LLPS process was further characterized using eq.\ref{eq:g2two}.
%The outcome is similar to the 2D scattering pattern, which we obtain as the raw data for the XPCS-USAXS experiment. Following the same experimental data procedure (equation \ref{eq:g2two}), we investigated the dynamics of the simulated LLPS process. 

\begin{figure}

 \includegraphics[width=6cm]{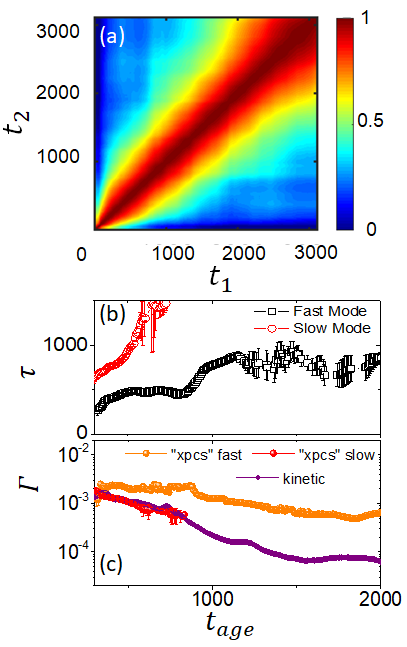}
 \caption{Kinetics and dynamics of LLPS simulated using CH equation: \textbf{(a)} Representative TTC for $T=0.05T_c$ with a total simulation time of $t_{age}=3105$. \textbf{(b)} Relaxation times of the two modes as a function of $t_{age}$. \textbf{(c)} Comparison of decay rates for the slow (coral red) and fast (orange) modes of dynamics with kinetics (purple) for simulated LLPS. Here $q_{simulation}$=22 pixel which is similar to the experimental $q$ = $4.4\mu m^{-1}$ (see Fig. S3 \textbf{(a)} and \textbf{(c)}). }
 \label{fgr:sim_dynamics}
\end{figure}

The main results of the simulations are presented in Fig. \ref{fgr:sim_dynamics}. They are highly dependent on the $q$-value. In order to qualitatively compare the simulation with experimental data, we focus on the similar $q$-region in comparison to the $q$ value of the early stage peak position of scattered intensity (SI Fig. S3). The total time of the simulation was estimated from the behavior of the scattering intensity for the chosen $q$ with time $t_{age}$. For the described parameters the simulated TTC reproduces the main features of the experimental TTC (compare Fig. \ref{fgr:sim_dynamics} (a) and Fig. \ref{fgr:TTC} (b)). TTCs from the simulation also show a square feature as observed in the experimental TTC.

The $g_2$ functions from simulations also contain two relaxations, which demonstrates the existence of the two modes of dynamics in a general case. The relaxation time for the slow mode linearly increases with time (SI Fig.\ref{fgr:sim_dynamics} (b)). For the fast mode, it increases with a modulation before reaching a steady state. The shape parameters $\gamma>1$ show behavior similar to the experimental results  (Fig. S2 (b)). Fig. \ref{fgr:sim_dynamics} (c) compares the relaxation rates. The relaxation rate of the slow mode obtained from correlation analysis exhibits behavior similar to the kinetics. Importantly, a simple classical model used for the current simulation of the phase transition demonstrates that the dynamics is much faster than the kinetics. This generalizes the result of time scale separation and widens its application from the specific one to a broad range of LLPS systems.  

It is worth noting that the simulation does not consider thermal motion, the limited heating rate, and the viscoelastic properties of the real system. This may be the reason that the early stage of simulation results do not show the exponential growth of $\tau$ (SI Fig. S5). The early stage of LLPS is sensitive to the thermal motion \cite{COOK1970}. It may result in the loss of the spatial correlations, which influences the visibility of the kinetic relaxation rate (being much slower than thermal motion effects) on the TTC. Thus, the dynamics is dominating in the experimental TTC. In the simulated TTC this effect is not included, and the kinetic relaxation rate can be seen in the early stage. The influence of the thermal motion decreases with the increase of the domain sizes, resulting in the appearance of the kinetic relaxation rate in the TTC (square feature) and finally becomes negligible in the coarsening stage. 

In summary, we have studied the non-equilibrium dynamics and structural kinetics of a protein solution undergoing a liquid-liquid phase separation using USAXS-XPCS. The results show that microscopic dynamics is up to 40 times in the early stage and still at least one magnitude faster in the late coarsening stage than the structural relaxation rate. Thus, these two components could be decoupled. In the early stage, the microscopic dynamics is not Brownian dynamics of the proteins, which would have required relaxation time $\propto q^{-2}$ and shape parameter $\gamma \sim 1$. Instead, this dynamics can be well described using a hyper-diffusive ballistic motion, i.e. the relaxation time $\propto q^{-1}$ and shape parameter $\gamma \sim 2$. Furthermore, the relaxation time of the dynamics exponentially increases with time in this early stage. In the late stage, where the domain coarsening is the main process, the relaxation time increase with time following a power law. 

Cahn-Hilliard simulations support the experimental results and broaden the conclusions to LLPS phenomena in general, making our results applicable for other soft matter and biological systems undergoing phase transition, where kinetics and dynamics are intertwined. For example, LLPS in living cells leads to various types of condensates with distinct dynamic properties, such as dense liquid, gel, and glass-like states \cite{Shin2017}. These distinct dynamic properties influence not only the kinetics of LLPS, but also the material transport and their biological functions in the crowded cellular environment. Finally, we emphasize that this work can be extended to different length and time scales using XPCS in SAXS mode and the fast development of X-ray free-electron laser (XFEL) facilities, so the dynamic behavior ranging from single protein to the domain coarsening could be covered.

This work was supported by the DFG and BMBF.  A.R acknowledges the Studienstiftung des deutschen Volkes for a personal fellowship, N.B. acknowledges the Alexander von Humboldt-Stiftung for a postdoctoral research fellowship and C.G. acknowledges BMBF (05K19PS1 and 05K20PSA) for financial support. 

%\balance


%%%REFERENCES%%%
\bibliography{apssamp}
\end{document}

\preprint{APS/123-QED}

\title{Supplementary Information
\\Interplay between Kinetics and Dynamics of Liquid-Liquid Phase Separation in a Protein Solution Revealed by Coherent X-ray Spectroscopy}% Force line breaks with \\

\author{Anastasia Ragulskaya}
\email{anastasia.ragulskaya@uni-tuebingen.de}
\affiliation{Institut\,f\"{u}r\,Angewandte\,Physik,\,Universit\"{a}t\,T\"{u}bingen,\,Auf\,der\,Morgenstelle\,10,\,72076\,T\"{u}bingen,\,Germany}
\author{Nafisa Begam}
\affiliation{Institut\,f\"{u}r\,Angewandte\,Physik,\,Universit\"{a}t\,T\"{u}bingen,\,Auf\,der\,Morgenstelle\,10,\,72076\,T\"{u}bingen,\,Germany}
\author{Anita Girelli}
\affiliation{Institut\,f\"{u}r\,Angewandte\,Physik,\,Universit\"{a}t\,T\"{u}bingen,\,Auf\,der\,Morgenstelle\,10,\,72076\,T\"{u}bingen,\,Germany}
\author{Hendrik Rahmann}
\affiliation{Department Physik, Universit\"{a}t Siegen, Emmy-Noether-Campus, Walter-Flex-Strasse 3, 57072 Siegen, Germany}
\author{Mario Reiser}
\affiliation{Department Physik, Universit\"{a}t Siegen, Emmy-Noether-Campus, Walter-Flex-Strasse 3, 57072 Siegen, Germany}
\affiliation{European X-ray Free-Electron Laser GmbH, Holzkoppel 4, 22869 Schenefeld, Germany}
\author{Fabian Westermeier}
\affiliation{Deutsches Elektronen-Synchrotron (DESY), Notkestrasse 85, 22607 Hamburg, Germany}
\author{Michael Sprung}
\affiliation{Deutsches Elektronen-Synchrotron (DESY), Notkestrasse 85, 22607 Hamburg, Germany}
\author{Fajun Zhang}
\email{fajun.zhang@uni-tuebingen.de}
\affiliation{Institut\,f\"{u}r\,Angewandte\,Physik,\,Universit\"{a}t\,T\"{u}bingen,\,Auf\,der\,Morgenstelle\,10,\,72076\,T\"{u}bingen,\,Germany}
\author{Christian Gutt}
\email{christian.gutt@uni-siegen.de}
\affiliation{Department Physik, Universit\"{a}t Siegen, Emmy-Noether-Campus, Walter-Flex-Strasse 3, 57072 Siegen, Germany}
\author{Frank Schreiber}
\email{frank.schreiber@uni-tuebingen.de}
\affiliation{Institut\,f\"{u}r\,Angewandte\,Physik,\,Universit\"{a}t\,T\"{u}bingen,\,Auf\,der\,Morgenstelle\,10,\,72076\,T\"{u}bingen,\,Germany}

\date{\today}% It is always \today, today,
\maketitle
\section{Materials and sample preparation}
BSA and $\textrm{YCl}_3$ were purchased from Sigma-Aldrich. Protein and salt were separately dissolved in degassed Milli-Q water (Merck Millipore, the resistivity of 18.2 M$\Omega\cdot$cm). The concentration of protein stock solution was measured using UV absorption \cite{Hay,DaVela16}. Then, the stock solutions of protein and salt were mixed at 21 $^\circ$C with adjusting of the concentrations by dilution with pure Milli-Q water, so that the final protein concentration was of 175 mg mL$^{-1}$ and salt concentration was of 42 mM (two-phase regime of the phase diagram \cite{Matsarskaia_2016_JPhysChemB,DaVela16}). Finally, the obtained solution was stabilized, centrifuged and only the dense phase was taken for further experiment.
\section{X-ray photon correlation spectroscopy (XPCS)}

XPCS experiments were performed in the USAXS mode at the PETRA III beamline P10 (DESY, Hamburg, Germany) at an incident X-ray energy of 8.54 keV ($\lambda$ = 1.452 \r{A}) and beam size of  $100$ $\mu m$ $\times$ 100 $\mu m$. The sample to detector distance was 21.2 m, which corresponds to the $q$ range of $3.2\times10^{-3}$ nm$^{-1}$ to $3.35\times10^{-2}$ nm$^{-1}$, where $q=4\pi /\lambda\cdot \sin \theta$ and $2\theta$ is the scattering angle. The data were collected by an EIGER X 4M CCD detector from Dectris with $75$ $\mu m$ $\times$ 75 $\mu m$ pixel size.  The sample was filled into capillaries of 1.5 mm in diameter and first equilibrated at 10 $^\circ$C for 10 minutes. Then, it was heated by the Linkam heating stage to 40 $^\circ$C with a heating rate of 150  $^\circ$C/min and measured for 312 s and 60 s from the beginning of the heating. After the measurement, the solution was cooled back to 10  $^\circ$C. Each measurement was taken on a fresh sample spot separated by larger than the beam size distance from the previous one to avoid the beam damage due to the long exposures. The number of frames, exposure time, and delay time (SI Table \ref{table:parameters}) were calculated based on the beam damage test, which was done preliminary to all measurements. The sample was measured with different attenuators at 10 $^\circ$C, so that we checked the behavior of the sample under the beam in one phase regime. The time of survival of the sample was considered as time with no change in USAXS profile and simultaneous noise in TTC and equaled to 60 s under the flux $1.42\times 10^9$ ph/s, which was taken into account for all further measurements. The temperature of the sample was recorded every 2 seconds. First 60 s and 312 s experiments were done under different exposure parameters (while under the beam damage threshold). However, the overlapping time period demonstrated the same kinetics and dynamics, which is additional evidence for selecting the beam parameters that do not affect the sample.

A series of 2D speckle patterns were obtained, which contains the dynamic information of the investigated sample. The same 2D scattering patterns were used for the investigation of the kinetics. To obtain the USAXS profile $\it{I(q)}$ for each frame (time), the scattering intensity was azimuthally averaged for different radii and the profile collected before the heat was used as the background.

\section{Cahn-Hilliard simulation}
A system undergoing spinodal decomposition can be described by the Cahn-Hilliard equation  \cite{Cahn1958,Cahn1959,Sappelt,Sciortino} (CH):

\begin{equation} \label{eq:CH}
\frac{\partial u(\mathbf{r}, t)}{\partial t}=\nabla\left[m(u) \nabla\left(-\frac{T_c-T}{T_c}u+u^{3}-\nabla^{2} u\right)\right],
\end{equation}
where $u(\textbf{r},t)$ - rescaled time-dependent local concentration, $m(u)$ - mobility function and $T_c$ - critical temperature. The $u(\textbf{r},t)$  can take values from -1 to 1. Positive values correspond to the dense phase and negative values to the diluted phase. For classical SD, the mobility function $m(u)$ is constant with time (set to 1). 

Equation \ref{eq:CH} was solved numerically in 2D using the second-order implicit Crank-Nicolson scheme \cite{Sappelt2}. To linearize the problem, the Newton-Raphson method was used \cite{Sciortino}. A space was divided in a matrix with size $N\times N$=$256\times256$ pixels. The initial concentration for each point in the space was defined as random noise fluctuations with amplitude $0.05$ around $u_{0}=0.4$, which corresponds to the dense phase (similar to the experiment), so that these noise fluctuations have a uniform distribution. The time step was $0.09$, the total number of steps was $34500$, and $T$ was set to $0.05 T_c$.

\begin{table*}[h]
  \caption{\ Experimental parameters}
  \label{table:parameters}
  \begin{ruledtabular}
  \begin{tabular}{ccccc}
   
    Duration (s) &  Number of frames &Exposure (s) & Delay (s) & $\sim$ Time of average (s)\\
    \hline
    60 & 7500& 0.008 & 0 & 0.08 \\
    312 & 3000& 0.004 & 0.1 & 1 \\

  \end{tabular}
  \end{ruledtabular}
\end{table*}

\begin{figure*}[h]
 \centering
 \includegraphics[height=7cm]{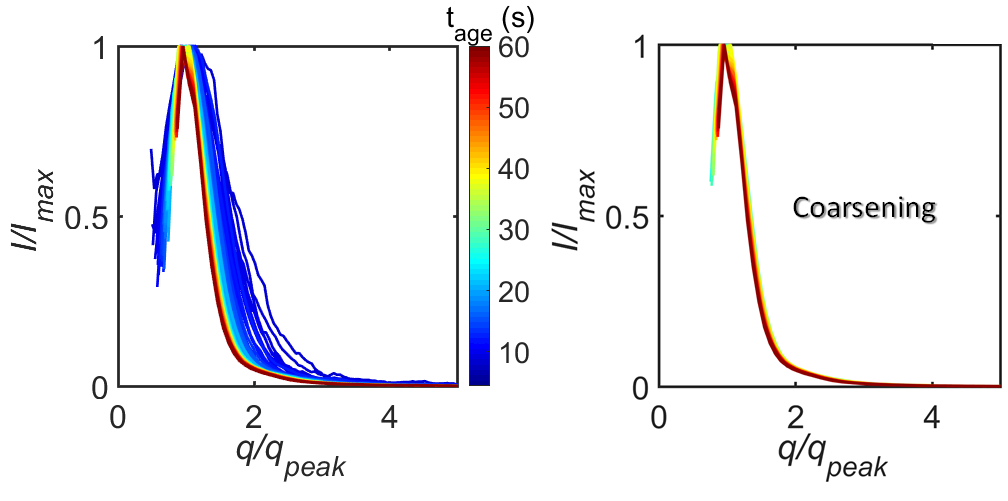}
 \caption{ Application of scaling to the scattered intensity curve for different experimental times $t_{age}$. The scaling law is not valid for the early stage of SD ($\sim$ 24 s): the normalised curves do not lie one on another (blue curves on the left graph). While the coarsening stage demonstrates compliance with the scaling law (right graph).
 }
 \label{fgr:scaling_law}
\end{figure*}

\begin{figure*}[h]
\centering
\includegraphics[height=7cm]{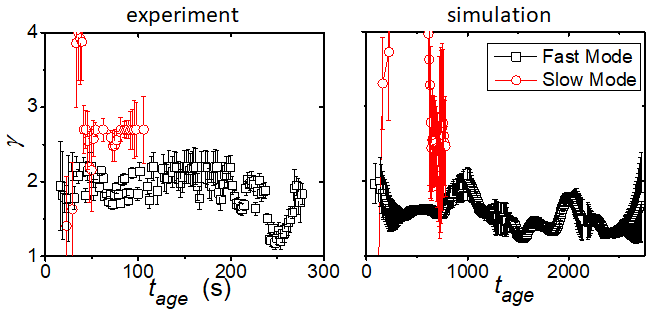}
\caption{Shape parameter $\gamma$ evolution with $t_{age}$ for fast (black squares)  and slow modes (red circles) of dynamics for experiment and Cahn-Hilliard simulation. 
}
\label{fgr:KWW}
\end{figure*}

\begin{figure*}[h]
 \centering
 \includegraphics[height=9cm]{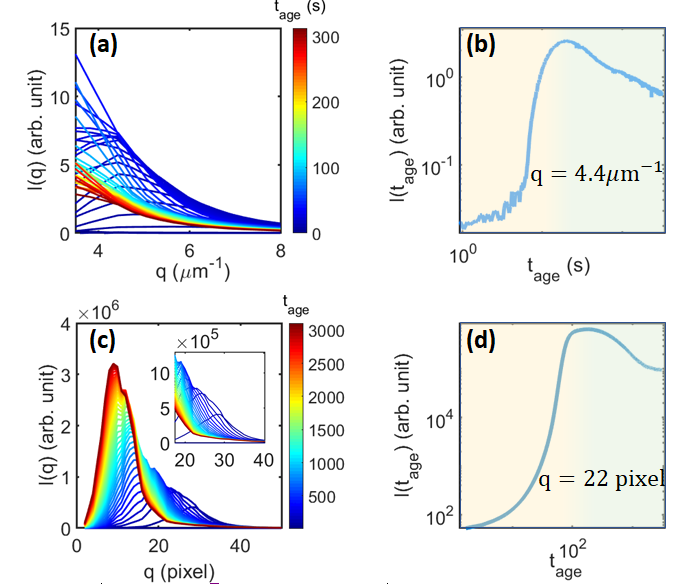}
 \caption{\textbf{(a)} Intensity profiles obtained by USAXS experiment at different experimental times $t_{age}$  for $T=40^\circ C$. (b) Intensity at $q=4.4\mu m^{-1}$ for $T=40^\circ C$. (c) Simulated intensity profiles obtained from Fourier transform of the real space image for $T=0.05T_c$. In the right corner there is a zoom for $q=[20,40] pixel$ range. (d) Intensity for simulated data for $T=0.05T_c$ at $q=22 pixel$. It is clearly seen that $q=22 pixel$ of the simulated intensity profile and $q=4.4\mu m^{-1}$ of the experimental profile have the same behavior (compare (d) with (b)) and these $q$-values are similar relative to the peak position of scattered intensity in the beginning (compare picture in the right corner of (c) with (a)). }
 \label{fgr:supporting_kinetics}
\end{figure*}

\begin{figure*}
\centering
\includegraphics[height=9cm]{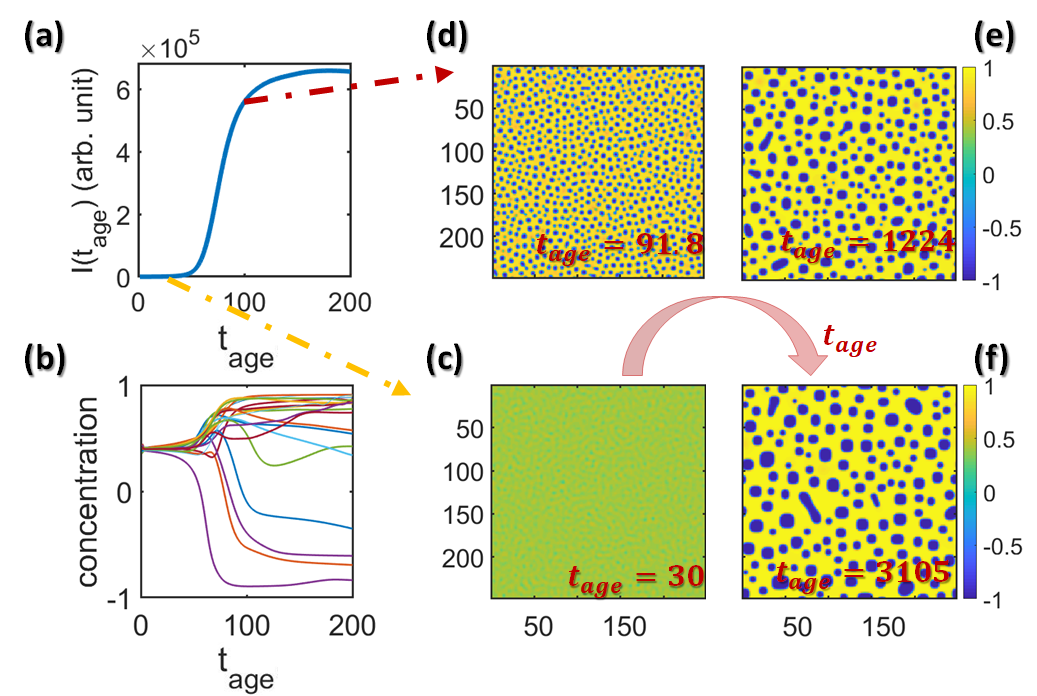}
\caption{(a) Behavior of scattered intensity with $t_{age}$ for the early stage of LLPS ($t_{age}$<200) for $T=0.05T_c$ for simulated data for the same $q=22 pixel$ for which Fig. 3 in the manuscript was presented. (b) Change of concentration for different random points (one point - one line) in the real space with time $t_{age}$ during early stage of LLPS. Time of significant change in concentration functions has similar value to the time of jump of structure factor on (a). (c)- (f) Simulated real space pictures of process of LLPS for different $t_{age}$ in the 248×248 pixels square. Colormap represents normalized concentration u(m,n, $t_{age}$) values. Yellow and Red dash dot lines show correspondence of the structure factor value and state of the system in the real space at the same time $t_{age}$.}
\label{fgr:SM concentration}
\end{figure*}

\begin{figure*}
\centering
\includegraphics[height=9cm]{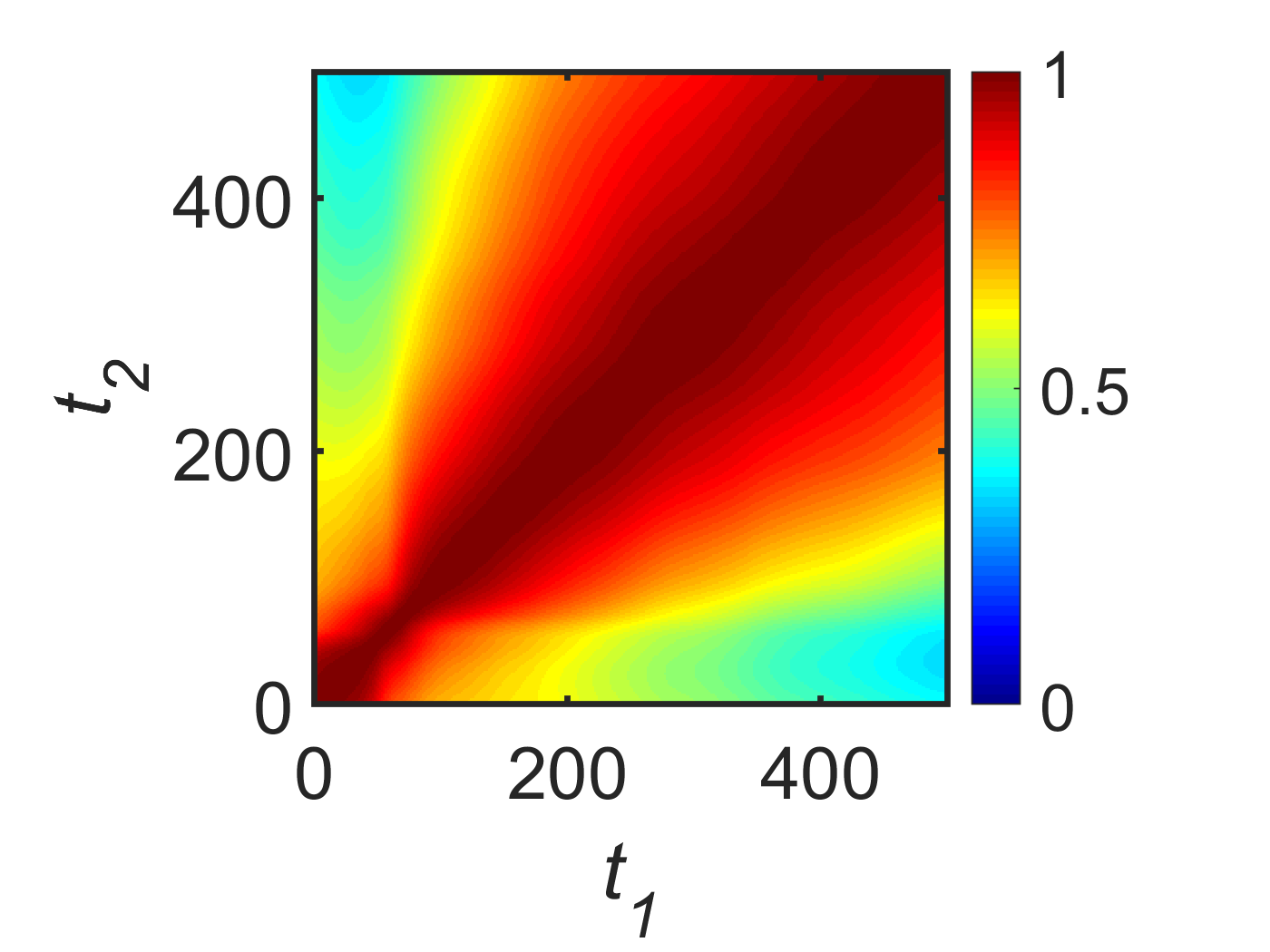}
\caption{Initial part of simulated ttc. It can be clearly seen that the simulation does not reproduce quick exponential increase in the early stage of SD detected at the experimental ttc. Furthermore, the $g_2$ cuts at first $\sim 300$ frames does not decrease to 0, which makes fit difficult.}
\label{fgr:simulation_ttc_initial}
\end{figure*}

\bibliography{apssamp}

%\nocite{*}

% Produces the bibliography via BibTeX.